\documentclass[final,3p,times,numbers,sort,compress]{elsarticle}

\usepackage{amsmath}
\usepackage{amsfonts}
\usepackage{amsmath}
\usepackage{amssymb}
\usepackage{amsthm}
\usepackage[utf8]{inputenc}
\usepackage{bm}
\usepackage{graphicx}
\usepackage{xcolor}

\usepackage[breaklinks=true,colorlinks=true,linkcolor=blue,urlcolor=blue,citecolor=blue]{hyperref}

\newcommand{\defeq}{\mathrel{\mathop:}=}
\newcommand{\U}{\mathcal{U}}
\renewcommand{\H}{\mathcal{H}}
\newcommand{\D}{\mathcal{D}}
\newcommand{\K}{\mathcal{K}}

\begin{document}

\begin{frontmatter}

\title{Superstatistics as the thermodynamic limit of driven classical systems}

\author[cchen,unab]{Sergio Davis\corref{cor1}}
\ead{sergio.davis@cchen.cl}
\address[cchen]{Research Center on the Intersection in Plasma Physics, Matter and Complexity, P$^2$mc,\\ Comisión Chilena de Energía Nuclear, Casilla 188-D, Santiago, Chile}
\address[unab]{Departamento de F\'isica, Facultad de Ciencias Exactas, Universidad Andres Bello,\\ Sazi\'e 2212, piso 7, 8370136, Santiago, Chile.}

\author[unab]{Claudia Loyola}
\author[unab]{Carlos Femenías}
\author[unab]{Joaquín Peralta}

\begin{abstract}
Superstatistics is an elegant framework for the description of steady-state thermodynamics, mostly used for systems with long-range interactions such as plasmas.
In this work, we show that the potential energy distribution of a classical system under externally imposed energy fluctuations can also be described by superstatistics in 
the thermodynamic limit. As an example, we apply this formalism to the thermodynamics of a finite Lennard-Jones crystal with constant microcanonical heat capacity driven by 
sinusoidal energy oscillations. Our results show that molecular dynamics simulations of the Lennard-Jones crystal are in agreement with the provided theoretical predictions.
\end{abstract}

\end{frontmatter}

\section{Introduction}

The thermodynamics of finite systems has gained attention recently, mostly in chemical physics~\cite{Dixit2013, Dixit2015} and condensed matter physics~\cite{Herron2021} in connection 
with the non-equilibrium generalization of Boltzmann-Gibbs statistics known as superstatistics~\cite{Beck2003, Beck2004}. This is motivated by the well-known failures in the 
thermodynamical description of small systems that exhibit significant deviations from the canonical behavior, especially near and below the critical point~\cite{Herron2021, Ritort2007}. 
In the superstatistics theory, commonly employed to describe systems with long-range interactions such as plasmas, self-gravitating systems and other complex systems, equilibrium is 
represented by a constant temperature, while non-equilibrium steady states are characterized by a statistical distribution of temperatures.

The superstatistical framework is clear and concise, expressing the idea of temperature fluctuations in a manner fully compatible with probability theory~\cite{Sattin2006}. However, 
it has been shown~\cite{Davis2022, Farias2023} that the description of the components of an isolated system needs to fulfill strict requirements to be compatible with 
superstatistics. For instance, if we divide an isolated system into two regions, superstatistics cannot describe one region if the other region has positive microcanonical heat capacity. 

In this work, we illustrate the application of the superstatistical framework to describe a finite-size system having sinusoidal energy oscillations, a simple realization of a driven system. We show 
that superstatistics is in fact an accurate description when the superstatistical inverse temperature $\beta \defeq 1/(k_B T)$ is taken to be the microcanonical inverse temperature. The rest of the 
paper is organized as follows. Section~\ref{sec:microcanonical} presents a detailed review of the microcanonical distribution and the potential energy of the system. Sections \ref{sec:superstat} and 
\ref{sec:oscillations} present a detailed review of the superstatistical framework and the thermodynamic limit for the case of sinusoidal energy oscillations; additionally, the theoretical results for 
the total and potential energy distributions are presented. Section~\ref{sec:md} presents the molecular dynamics simulations for a Lennard-Jones (LJ) finite-size system used to test our theoretical 
results. Finally, we give some concluding remarks and discuss the scope of our work in Section~\ref{sec:con}.

\newpage
\section{Microcanonical distribution of potential energy}
\label{sec:microcanonical}

In order to describe the probability density of potential energy of a classical finite-size system, we first begin with the particular case of fixed total energy, i.e. a microcanonical system. We will 
consider a system of $N$ particles with Hamiltonian
\begin{equation}
\label{eq:ham}
\H(\bm{\Gamma}) = K(\bm{p}_1, \ldots, \bm{p}_N) + \Phi(\bm{r}_1,\ldots,\bm{r}_N),
\end{equation}
where $\bm{\Gamma} = (\bm{r}_1, \ldots, \bm{r}_N, \bm{p}_1, \ldots, \bm{p}_N)$ is a point in phase space,
\begin{equation}
K(\bm{p}_1,\ldots,\bm{p}_N) = \sum_{i=1}^N \frac{\bm{p}_i^2}{2m_i}
\end{equation}
is the kinetic energy and $\Phi(\bm R)$ the potential energy. If our system is isolated, then the energy is fixed at a value $E$ and we have
\begin{equation}
P(\bm{r}_1, \ldots, \bm{r}_N|E) = \frac{1}{\Omega(E)}\int d\bm{p}_1\ldots d\bm{p}_N\,\delta\big(E-K(\bm{p}_1,\ldots,\bm{p}_N)-\Phi(\bm{r}_1,\ldots,\bm{r}_N)\big) \\
 = \frac{\Omega_K\big(E-\Phi(\bm{r}_1, \ldots, \bm{r}_N)\big)}{\Omega(E)}
\end{equation}
where $\Omega_K$ is the kinetic density of states, given by
\begin{equation}
\label{eq:kindos}
\Omega_K(K) \defeq \int d\bm{p}_1\ldots d\bm{p}_N\,\delta\left(\sum_{i=1}^N \frac{\bm{p}_i^2}{2m_i} - K\right) = W\,\big[K\big]_+^{\frac{3N}{2}-1},
\end{equation}
with $W$ a constant and where the notation $[x]_+ = \max(0, x)$ is used. The microcanonical heat capacity is defined by
\begin{equation}
C_E \defeq \left(\frac{dE}{dT}\right)_E = -k_B\,\frac{\beta_\Omega(E)^2}{{\beta_\Omega}'(E)}
\end{equation}
where
\begin{equation}
\label{eq:betaom_def}
\beta_\Omega(E) \defeq \frac{1}{k_B T(E)} = \frac{\partial}{\partial E}\ln \Omega(E),
\end{equation}
is the microcanonical inverse temperature. In this work we will be considering a system where $C_E$ is a constant, let us say
\begin{equation}
\label{eq:const_CE}
C_E = \alpha k_B,
\end{equation}
where $\alpha$ is an extensive, dimensionless constant. In this case we must have $E = \alpha k_B T(E)$, therefore
\begin{equation}
\beta_\Omega(E) = \frac{\alpha}{E}
\end{equation}
and, by integration of \eqref{eq:betaom_def}, it follows that the full density of states $\Omega(E)$ must be of the form
\begin{equation}
\label{eq:dos}
\Omega(E) = \Omega_0\,E^\alpha.
\end{equation}

\noindent
Using this we can write
\begin{equation}
\label{eq:probphi_E}
P(\phi|E) = \int d\bm{R} \delta\big(\Phi(\bm R)-\phi\big)P(\bm{R}|E) = \frac{W}{\Omega(E)}\big[E-\phi\big]_+^{\frac{3N}{2}-1}\mathcal{D}(\phi),
\end{equation}
where $\D(\phi)$ is the configurational density of states, defined by
\begin{equation}
\D(\phi) \defeq \int d\bm{R}\delta\big(\Phi(\bm R)-\phi).
\end{equation}

The distribution in \eqref{eq:probphi_E} is the microcanonical distribution of potential energies introduced in the context of molecular simulation~\cite{Severin1978,Pearson1985,Ray1991}.
As we prove in \ref{sec:appendix}, the only configurational density of states compatible with $\Omega(E)$ as in \eqref{eq:dos} has the form
\begin{equation}
\label{eq:cdos}
\mathcal{D}(\phi) = D_0\,\phi^{\alpha-\frac{3N}{2}}.
\end{equation}

It is important to note that the exponent $\alpha-\frac{3N}{2}$ in \eqref{eq:cdos} is exact for all $N \geq 1$. Replacing \eqref{eq:cdos} we can impose normalization 
of the distribution to obtain
\begin{equation}
\label{eq:probphi_beta}
P(\phi|E) = \frac{\phi^{\alpha-\frac{3N}{2}}\big[E-\phi\big]_+^{\frac{3N}{2}-1}}{B\big(\frac{3N}{2}, \alpha+1-\frac{3N}{2}\big)E^\alpha},
\end{equation}
where $B(a, b)$ is the beta function. The most probable value of $\phi$ given $E$, denoted by $\phi^*(E)$, can be determined from the extremum condition
\begin{equation}
0 = \left[\frac{\partial}{\partial \phi}\ln P(\phi|E)\right]_{\phi = \phi^*(E)} = -\frac{3N-2}{2\big(E-\phi^*(E)\big)} + \frac{2\alpha-3N}{2\phi^*(E)}
\end{equation}
from which it follows that
\begin{equation}
\label{eq:phi_mode}
\phi^*(E) = \left[1-\frac{3N-2}{2(\alpha-1)}\right]E.
\end{equation}

In order to determine the moments of $\phi$ at constant $E$ we will make use of the conjugate variables theorem~\cite{Davis2012, Davis2016}. This identity is a consequence of 
the divergence theorem and relates expectations of derivatives with expectations involving logarithmic derivatives of the probability density. For a probability density $P(X|S)$ 
it has the form
\begin{equation}
\left<\frac{\partial \omega}{\partial X}\right>_S + \left<\omega\frac{\partial}{\partial X}\ln P(X|S)\right>_S = 0,
\end{equation}
where $\omega(X)$ is an arbitrary, differentiable function of the variable $X$. In our case we have
\begin{equation}
\left<\frac{\partial \omega}{\partial \phi}\right>_E = \left<\omega\left[\frac{3N-2}{2(E-\phi)} - \frac{2\alpha-3N}{2\phi}\right]\right>_E
\end{equation}
with $\omega(\phi)$ an arbitrary, differentiable function of $\phi$. Using $\omega(\phi) = \phi^m (E-\phi)$ and after some algebra, we have the recurrence relation
\begin{equation}
(\alpha + m)\big<\phi^m\big>_E = \left(\alpha + m - \frac{3N}{2}\right)E\big<\phi^{m-1}\big>_E
\end{equation}
which has solution
\begin{equation}
\label{eq:moments_E}
\big<\phi^m\big>_E = E^m \prod_{k=1}^m \left(\frac{\alpha + k - \frac{3N}{2}}{\alpha + k}\right)
= \frac{\Gamma(\alpha+1)\Gamma(\alpha+m+1-\frac{3N}{2})}{\Gamma(\alpha+m+1)\Gamma(\alpha+1-\frac{3N}{2})}\,E^m.
\end{equation}

\noindent
From here, using $m = 1$ we obtain the mean potential energy as
\begin{equation}
\label{eq:mean_phi_E}
\big<\phi\big>_E = \left[1-\frac{3N}{2(\alpha+1)}\right]E,
\end{equation}
and, as expected, $\big<\phi\big>_E \approx \phi^*(E)$ for $N \gg 1$ and $\alpha \gg 1$. Similarly, using $m = 2$ we obtain the relative variance
\begin{equation}
\label{eq:relvar_E}
\frac{\big<(\delta \phi)^2\big>_E}{\big<\phi\big>_E^2} = \frac{3N}{2(\alpha+2)\big(\alpha-3N/2+1\big)}.
\end{equation}

The formula in \eqref{eq:relvar_E} provides a finite-size version of the celebrated Lebowitz-Percus-Verlet~\cite{Lebowitz1967} formula, used to calculate the heat capacity 
in atomistic computer simulations. Clearly, in the thermodynamic limit, $N \rightarrow \infty$ and $\alpha \rightarrow \infty$ so the relative variance of $\phi$ vanishes as 
$1/N$, as expected from equilibrium thermodynamics. It is also important to note, for the following sections, that
\begin{equation}
\left<\frac{3N-2}{2(E-\phi)}\right>_E = \beta_\Omega(E)
\end{equation}
for all sizes $N$, as can be deduced from the identity
\begin{equation}
0 = \left<\frac{\partial}{\partial E}\ln P(\phi|E)\right>_E = \left<\frac{3N-2}{2(E-\phi)} - \frac{\alpha}{E}\right>_E = \left<\frac{3N-2}{2(E-\phi)}\right>_E - \beta_\Omega(E),
\end{equation}
where we have replaced $P(\phi|E)$ as given in \eqref{eq:probphi_beta}.

\section{The superstatistical approximation in the thermodynamic limit}
\label{sec:superstat}

In this section we will generalize the microcanonical description of the previous section towards a steady state where energy does fluctuate, and formulate such a description in 
connection with the framework of superstatistics. Superstatistics~\cite{Beck2003,Beck2004} is regarded as an elegant proposal among those that aim to describe non-equilibrium systems 
in steady states. It promotes the constant inverse temperature $\beta = 1/(k_B T)$ of the canonical ensemble to a random variable, having a joint distribution
\begin{equation}
\label{eq:super_joint}
P(\bm{\Gamma}, \beta|S) = P(\bm{\Gamma}|\beta)P(\beta|S) = \left[\frac{\exp\big(-\beta \H(\bm \Gamma)\big)}{Z(\beta)}\right]P(\beta|S),
\end{equation}
with the microstates $\bm \Gamma$. The marginal distribution of microstates is then given by
\begin{equation}
P(\bm{\Gamma}|S) = \int_0^\infty d\beta P(\bm{\Gamma}|\beta)P(\beta|S),
\end{equation}
where we have integrated out the variable $\beta$. Replacing \eqref{eq:super_joint} we have that $P(\bm \Gamma|S)$ is a superposition of canonical models that considers every possible 
value of $\beta$ weighted by its probability density. That is, we can write
\begin{equation}
\label{eq:super}
P(\bm{\Gamma}|S) = \int_0^\infty d\beta P(\beta|S)\left[\frac{\exp\big(-\beta \H(\bm \Gamma)\big)}{Z(\beta)}\right].
\end{equation}

Therefore, our goal is to write a general ensemble $P(\bm \Gamma|S)$ with an arbitrary distribution of energy $P(E|S)$ as an instance of \eqref{eq:super} under a suitable definition of 
fluctuating temperature $\beta$ and its distribution $P(\beta|S)$. The ensemble function for $\phi$ given $E$ is
\begin{equation}
\rho(\phi; E) = \frac{W}{\Omega(E)}\big[E-\phi\big]_+^{\frac{3N}{2}-1}
\end{equation}
such that $P(\phi|E) = \rho(\phi; E)\D(\phi)$ agrees with \eqref{eq:probphi_E}. If we lift the constraint of constant total energy, using the marginalization rule 
we can write
\begin{equation}
P(\phi|S) = \int_0^\infty dE P(E|S)P(\phi|E),
\end{equation}
which implies
\begin{equation}
\label{eq:rho_mixture}
\rho(\phi; S) = \int_0^\infty dE P(E|S)\rho(\phi; E).
\end{equation}

In the following, we will show that $\rho(\phi; S)$ in \eqref{eq:rho_mixture} reduces for $N \gg 1$ to a superstatistical ensemble function. First, we note that the 
factor 
\begin{equation}
\mathcal{M}(\phi; E) \defeq \big[E-\phi\big]_+^{\frac{3N}{2}-1}
\end{equation}
can be rewritten as
\begin{equation}
\label{eq:factor_M}
\mathcal{M}(\phi; E) = (E-\phi_E)^{\frac{3N}{2}-1}\Big[1 - \frac{\phi-\phi_E}{E-\phi_E}\Big]_+^{\frac{3N}{2}-1}
= (E-\phi_E)^{\frac{3N}{2}-1}\Big[1-(1-q)\beta_E\,\phi\Big]_+^{\frac{1}{1-q}}
\end{equation}
where we have introduced a reference potential energy $\phi_E$, to be determined later, together with the quantities
\begin{subequations}
\begin{align}
q & \defeq 1 - \frac{2}{3N-2}, \\
\beta_E & \defeq \frac{3N-2}{2(E-\phi_E)}.
\end{align}
\end{subequations}

\noindent
We can approximate the factor $\mathcal{M}$ in \eqref{eq:factor_M} for $N \gg 1$ as
\begin{equation}
\mathcal{M}(\phi; E) \approx (E-\phi_E)^{\frac{3N}{2}}\exp(-\beta_E\,\phi),
\end{equation}
so that normalization of \eqref{eq:probphi_E} implies
\begin{equation}
\label{eq:dos_approx}
\Omega(E) = \int_0^\infty d\phi\,W\mathcal{M}(\phi; E)\D(\phi) \approx W(E-\phi_E)^{\frac{3N}{2}}\,Z(\beta_E).
\end{equation}
where $Z(\beta)$ is the configurational partition function, defined by
\begin{equation}
\label{eq:part}
Z(\beta) \defeq \int_0^\infty d\phi \D(\phi)\exp(-\beta \phi) = \frac{\Omega_0\Gamma(\alpha+1)}{W\,\Gamma\big(\frac{3N}{2}\big)}\beta^{\frac{3N}{2}-\alpha-1}.
\end{equation}

\noindent
Furthermore, replacing the approximations for $\mathcal{M}(\phi; E)$ and $\Omega(E)$ into \eqref{eq:probphi_E}, we see that
\begin{equation}
\rho(\phi; E) \approx \frac{W}{\Omega(E)}\big(E-\phi_E\big)^{\frac{3N}{2}} \exp(-\beta_E\,\phi)
\approx \left[\frac{\exp(-\beta\,\phi)}{Z(\beta)}\right]_{\beta = \beta_E},
\end{equation}
that is, the ensemble function given $E$ is that of a $q$-canonical distribution, as first shown by Naudts \emph{et al}~\cite{Naudts2009}, such that it reduces to a canonical ensemble at inverse 
temperature $\beta_E$ in the thermodynamic limit. In particular, the condition that for all energies $E$,
\begin{equation}
\label{eq:fix_phiE}
\big<\phi\big>_E = E\left(1-\frac{3N}{2(\alpha+1)}\right) \approx \big<\phi\big>_{\beta = \beta_E}
\end{equation}
with $\big<\phi\big>_\beta$ the canonical expectation of the potential energy, allows us to fix the value of $\phi_E$. We have
\begin{equation}
\big<\phi\big>_\beta = -\frac{\partial}{\partial \beta}\ln Z(\beta) = \frac{\alpha R}{\beta}
\end{equation}
where $R$ is a constant defined by
\begin{equation}
\label{eq:Rdef}
R \defeq 1 - \frac{3N}{2\alpha},
\end{equation}
therefore for $\alpha \gg 1$ the condition in \eqref{eq:fix_phiE} reduces to
\begin{equation}
E = \frac{2\alpha}{3N}(E-\phi_E)
\end{equation}
with solution $\phi_E = RE$. We see that $\phi_E$ coincides with the most probable potential energy $\phi^*(E)$ in \eqref{eq:phi_mode} for $N \gg 1$, and is such that 
\begin{equation}
\beta_E = \frac{\alpha}{E} = \beta_\Omega(E),
\end{equation}
thus there is no ambiguity in the value of inverse temperature. In summary, we have shown that \eqref{eq:rho_mixture} can be approximated in the thermodynamic limit 
by the superstatistical ensemble function
\begin{equation}
\label{eq:rho_super_approx}
\rho(\phi; S) \approx \int_0^\infty d\beta P(\beta|S)\left[\frac{\exp(-\beta \phi)}{Z(\beta)}\right]
\end{equation}
with a distribution of inverse temperatures given by
\begin{equation}
\label{eq:prob_beta}
P(\beta|S) = \int_0^\infty dE\,P(E|S)\delta\big(\beta_\Omega(E)-\beta\big).
\end{equation}

\noindent
Replacing \eqref{eq:part} in \eqref{eq:rho_super_approx} and imposing normalization we finally obtain
\begin{equation}
\label{eq:probphi_super}
P(\phi|S) \approx \frac{1}{\Gamma\big(\alpha+1-\frac{3N}{2}\big)}\int_0^\infty d\beta P(\beta|S)\beta\cdot(\beta\phi)^{\alpha-\frac{3N}{2}}\exp(-\beta \phi).
\end{equation}

\section{Energy oscillations}
\label{sec:oscillations}

\noindent
To study the effect of energy oscillations in a finite-size system, for simplicity, we will consider that our system has a time-dependent total energy given by
\begin{equation}
\label{eq:sinusoidal}
E(t) = E_0 + A\sin\,(\omega t)
\end{equation}
so that $E(t) \in \big[E_0-A, E_0+A\big]$. The probability of observing a value $E$ is obtained from the time average
\begin{equation}
\label{eq:probE_dyn}
P(E|A, E_0, \omega) = \lim_{\tau \rightarrow \infty} \frac{1}{\tau}\int_0^\tau dt\,\delta\big(E-E(t)\big)
= \lim_{\tau \rightarrow \infty} \frac{1}{\tau}\int_0^\tau dt \sum_{t_0} \frac{\delta\big(t-t_0\big)}{|E'(t_0)|}
\end{equation}
where the times $t_0$ are such that
\begin{equation}
\sin\,(\omega t_0) = \frac{E-E_0}{A}.
\end{equation}

\noindent
By replacing
\begin{equation}
E'(t_0) = A\omega\cos\,(\omega t_0) = \omega\sqrt{A^2-(E-E_0)^2}
\end{equation}
in \eqref{eq:probE_dyn} we obtain the probability density
\begin{equation}
\label{eq:probE}
P(E|A, E_0) = \frac{1}{\pi}\frac{\Theta\big(A-|E-E_0|\big)}{\sqrt{A^2-(E-E_0)^2}}
= \frac{1}{\pi}\frac{\Theta\big(A-|E-E_0|\big)}{\sqrt{E-(E_0-A)}\sqrt{(E_0+A)-E}},
\end{equation}
which gives the energy fluctuations associated with the oscillation. Note that this distribution does not depend on the frequency of oscillation, only on the 
amplitude $A$ and the reference energy $E_0$.

\noindent
The superstatistical distribution of inverse temperatures can be directly computed from \eqref{eq:probE} and \eqref{eq:prob_beta}, yielding
\begin{equation}
\label{eq:probbeta}
P(\beta|A, E_0) = \frac{1}{\pi \beta_0\sqrt{\gamma^2 (\beta/\beta_0)^4 - (\beta/\beta_0)^2 (1 - \beta/\beta_0)^2}}
\end{equation}
provided that 
\begin{equation}
\frac{1}{1+\gamma} < \frac{\beta}{\beta_0} < \frac{1}{1-\gamma},
\end{equation}
where we have defined a reference inverse temperature $\beta_0 \defeq \alpha/E_0$. The first and second moments of \eqref{eq:probbeta} are
\begin{equation}
\beta_S \defeq \big<\beta\big>_{A, E_0} = \frac{\beta_0}{\sqrt{1-\gamma^2}},
\end{equation}
and
\begin{equation}
\big<\beta^2\big>_{A, E_0} = \frac{(\beta_S)^2}{\sqrt{1-\gamma^2}},
\end{equation}
respectively, and from them we can compute the superstatistical relative variance of $\beta$ as
\begin{equation}
\label{eq:super_relvar}
u \defeq \frac{\big<(\delta \beta)^2\big>_{A, E_0}}{\big<\beta\big>_{A, E_0}^2} = \frac{1}{\sqrt{1-\gamma^2}}-1.
\end{equation}

Direct calculation of the energy moments from \eqref{eq:probE} is cumbersome, and we will again make use of the conjugate variables theorem.
For the case of the distribution $P(E|A, E_0)$ in \eqref{eq:probE} it reduces to
\begin{equation}
\left<\frac{\partial \omega}{\partial E}\right>_{A, E_0} = -\left<\omega\left[\frac{E-E_0}{A^2-(E-E_0)^2}\right]\right>_{A, E_0}.
\end{equation}

\noindent
Under the choice $\omega(E) = g(E)\big[A^2-(E-E_0)^2\big]$, where $g(E)$ is another arbitrary, differentiable function of $E$, we have
\begin{equation}
\label{eq:cvt}
\left<\frac{\partial g}{\partial E}\Big[A^2 - (E-E_0)^2\Big]\right>_{A, E_0} = \Big<g(E)(E-E_0)\Big>_{A,E_0}
\end{equation}

\noindent
Using $g(E) = 1$ gives the mean energy as
\begin{equation}
\big<E\big>_{A, E_0} = E_0,
\end{equation}
while from $g(E) = E-E_0$ we obtain the energy variance
\begin{equation}
\big<(\delta E)^2\big>_{A, E_0} = \frac{A^2}{2},
\end{equation}
as expected from \eqref{eq:sinusoidal}. A recurrence relation for the moments of $E$ can be obtained from \eqref{eq:cvt} under the choice $g(E) = E^m$, namely
\begin{equation}
\label{eq:recurr_E}
m\big(\gamma^2-1\big){E_0}^2\big<E^{m-1}\big>_{A, E_0} + (2m+1)E_0\big<E^m\big>_{A, E_0} = (m+1)\big<E^{m+1}\big>_{A, E_0}
\end{equation}
where we have introduced 
\begin{equation}
\gamma \defeq \frac{A}{E_0}.
\end{equation}

In order to solve it, it is more convenient to compute the moment generating function for the variable $E-E_0$, defined by
\begin{equation}
M_{E-E_0}(t; A, E_0) \defeq \big<\exp\big(t(E-E_0)\big)\big>_{A, E_0},
\end{equation}
which can be obtained from \eqref{eq:cvt} using $g(E) = \exp\big(t(E-E_0)\big)$. This leads to
\begin{equation}
A^2 t\,\Big<\exp\big(t\big[E-E_0\big]\big)\Big>_{A, E_0} - t\left<\exp(t\big[E-E_0\big])(E-E_0)^2\right>_{A, E_0} = \Big<\exp\big(t\big[E-E_0\big]\big)(E-E_0)\Big>_{A,E_0},
\end{equation}
which can be transformed into the second-order differential equation
\begin{equation}
A^2 t\,M_{E-E_0}(t; A, E_0) - t\frac{\partial^2}{\partial t^2}M_{E-E_0}(t; A, E_0) = \frac{\partial}{\partial t}M_{E-E_0}(t; A, E_0)
\end{equation}
with solution
\begin{equation}
\label{eq:mgf1}
M_{E-E_0}(t; A, E_0) = I_0(At)
\end{equation}
where $I_0$ is the modified Bessel function of the first kind. The moment generating function for $E$ follows directly from \eqref{eq:mgf1}, as
\begin{equation}
M_E(t; A, E_0) \defeq \big<\exp(tE)\big>_{A, E_0} = \exp(t\,E_0)M_{E-E_0}(t; A, E_0) = \exp(t\,E_0)I_0(At),
\end{equation}
and from it we can extract the moments
\begin{equation}
\label{eq:moments}
\big<E^m\big>_{A, E_0} = \phantom{.}_2F_1\Big(\frac{1-m}{2}, -\frac{m}{2}, 1; \gamma^2\Big){E_0}^m,
\end{equation}
in which $\phantom{.}_2F_1$ corresponds to the Gauss hypergeometric function, for $m \geq 0$. We will now compute the inverse temperature covariance $\U$, introduced in Ref.~\cite{Davis2022} and defined by
\begin{equation}
\U \defeq \big<(\delta \beta_\Omega)^2\big>_S + \big<{\beta_\Omega}'\big>_S,
\end{equation}
and which is relevant to determine the superstatistical regime. In superstatistics, $\U$ is equal to the variance of the superstatistical parameter 
$\beta$, so $\U \geq 0$ and from \eqref{eq:super_relvar} we expect that
\begin{equation}
\U \approx \big<(\delta \beta)^2\big>_{A, E_0} = (\beta_S)^2 \left[\frac{1}{\sqrt{1-\gamma^2}}-1\right].
\end{equation}
 
For the computation of $\U$ we need the expected values of $\big<E^{-1}\big>_{A, E_0}$ and $\big<E^{-2}\big>_{A, E_0}$, and because the moments in \eqref{eq:moments} are 
solutions of the recurrence relation in \eqref{eq:recurr_E}, they are also valid for $m < 0$. The expectation of $\beta_\Omega$ can be computed using $m = -1$ in \eqref{eq:moments}, giving
\begin{equation}
\big<\beta_\Omega\big>_{A, E_0} = \left<\frac{\alpha}{E}\right>_{A, E_0} = \frac{\beta_0}{\sqrt{1-\gamma^2}} = \beta_S.
\end{equation}
 
We see then that $\beta_S$ increases, i.e. the steady state temperature decreases, with $\gamma$. From this result for $\beta_S$, and using \eqref{eq:const_CE}, we can write the generalized equation of state
\begin{equation}
\big<E\big>_{A, E_0} = \frac{C_E}{\sqrt{1-\gamma^2}}\,T_S,
\end{equation}
where we have a generalized heat capacity
\begin{equation}
C_S \defeq \frac{C_E}{\sqrt{1-\gamma^2}} \geq C_E.
\end{equation}

\noindent
Using $m = -2$ in \eqref{eq:moments}, we readily have
\begin{equation}
\big<E^{-2}\big>_{A, E_0} = \frac{1}{{E_0}^2}\frac{1}{\big(\sqrt{1-\gamma^2}\big)^3},
\end{equation}
so we can write
\begin{subequations}
\begin{align}
\big<(\delta \beta_\Omega)^2\big>_{A, E_0} & = \alpha^2\big<E^{-2}\big>_{A, E_0} - {\beta_S}^2, \\
\big<{\beta_\Omega}'\big>_{A, E_0} & = -\alpha\big<E^{-2}\big>_{A, E_0}.
\end{align}
\end{subequations}

\noindent
After some algebra, we obtain
\begin{equation}
\U = u\,(\beta_S)^2
\end{equation}
with
\begin{equation}
\label{eq:upred}
u = \frac{\alpha-1}{\alpha\sqrt{1 - \gamma^2}} - 1.
\end{equation}
 
As expected, the reduced inverse temperature covariance increases with $\gamma$, as the system is driven further away from equilibrium. There is a critical value of $\gamma$, 
namely
\begin{equation}
\gamma_c \defeq \frac{\sqrt{2\alpha-1}}{\alpha}
\end{equation}
such that $\gamma > \gamma_c$ leads to $u > 0$, making the ensemble supercanonical. Note also that
\begin{equation}
\label{eq:upred_asympt}
\lim_{\alpha \rightarrow \infty} u = \frac{1}{\sqrt{1-\gamma^2}}-1 \geq 0,
\end{equation}
therefore in that limit we have agreement with \eqref{eq:super_relvar}. The threshold $\gamma_c$ goes to zero as $\alpha \rightarrow \infty$, consistent with the idea that, in the thermodynamic limit, 
the ensemble must be compatible with superstatistics, and therefore must have $u > 0$ for all $\gamma$. On the other hand, for vanishing oscillation amplitude we have
\begin{equation}
\lim_{\gamma \rightarrow 0} u = -\frac{1}{\alpha},
\end{equation}
recovering the microcanonical result. Alternatively, we can obtain $u$ and $\beta_S$ from 
\begin{equation}
\label{eq:beta_S_alt}
\beta_S = \big<b_\Omega\big>_{A, E_0}
\end{equation}
and
\begin{equation}
\label{eq:umeas}
u = \frac{\big<(\delta b_\Omega)^2\big>_{A, E_0} + \big<{b_\Omega}'\big>_{A, E_0}}{\big<b_\Omega\big>_{A, E_0}}
\end{equation}
where
\begin{equation}
b_\Omega(\phi) \defeq \frac{\partial}{\partial \phi}\ln \D(\phi) = \frac{2\alpha-3N}{2\phi}.
\end{equation}

The exact distribution of potential energies for any value of $A$ and any system size $N$ is given by marginalization of the energy variable $E$, that is, by the integral
\begin{equation}
P(\phi|A, E_0) = \int_0^\infty dE P(E|A, E_0)P(\phi|E).
\end{equation}

\noindent
Replacing \eqref{eq:probE} and \eqref{eq:probphi_beta} we can write
\begin{equation}
\label{eq:probphi_exact}
P(\phi|A, E_0) = \frac{\Big(\frac{\phi}{E_0}\Big)^{\alpha-\frac{3N}{2}}\K\Big(\frac{\phi}{E_0}; \gamma, \alpha, N\Big)}{E_0\,\pi\,B\big(\frac{3N}{2}, \alpha+1-\frac{3N}{2}\big)}
\end{equation}
where we have defined the special function
\begin{equation}
\K(z; \gamma, \alpha, N) \defeq \int_{\max\,(z, 1-\gamma)}^{1+\gamma} \frac{dx\,x^{-\alpha}(x-z)^{\frac{3N}{2}-1}}{\sqrt{x-(1-\gamma)}\sqrt{(1+\gamma)-x}}.
\end{equation}

\begin{figure}[t!]
\begin{center}
\includegraphics[height=0.25\textheight]{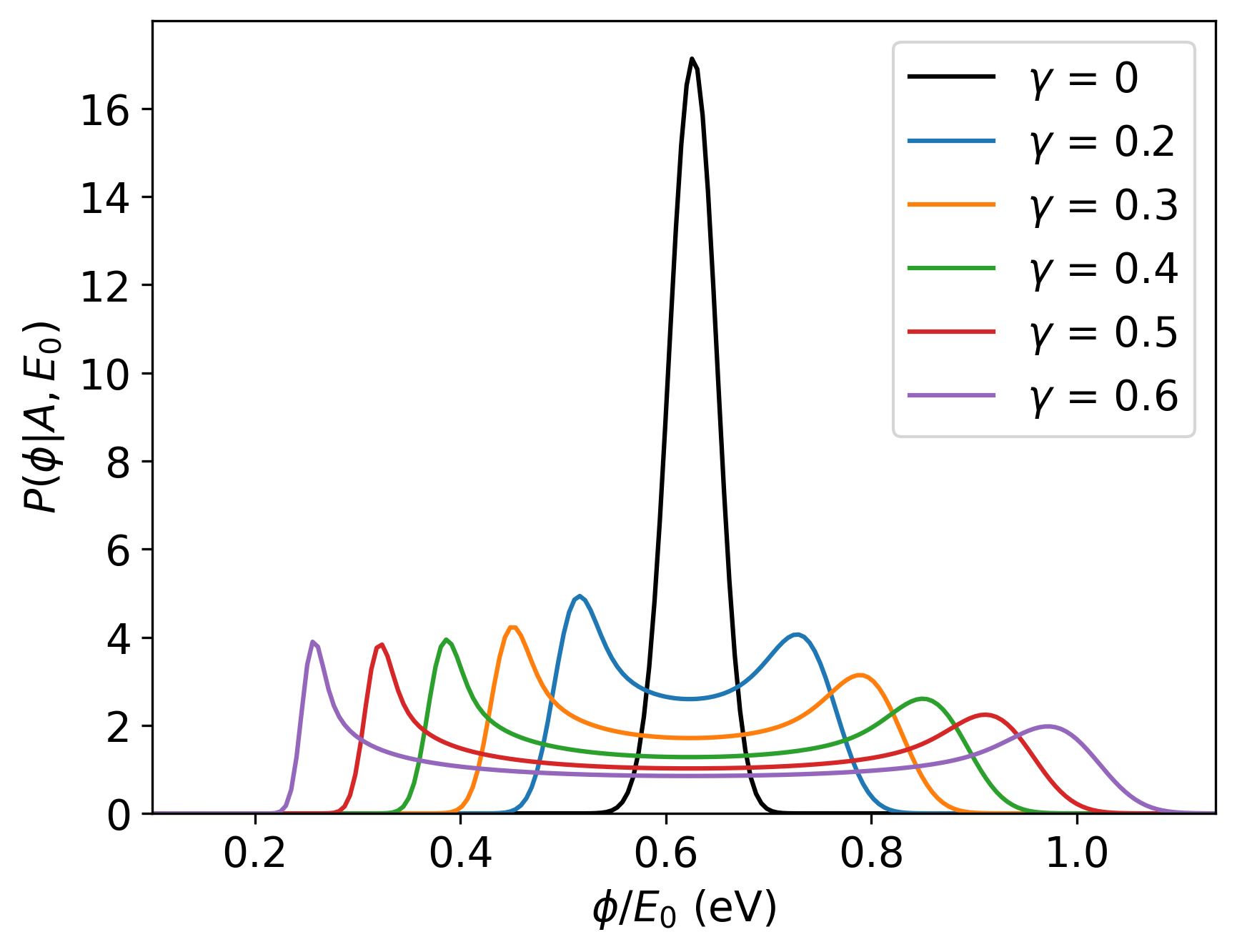}
\includegraphics[height=0.25\textheight]{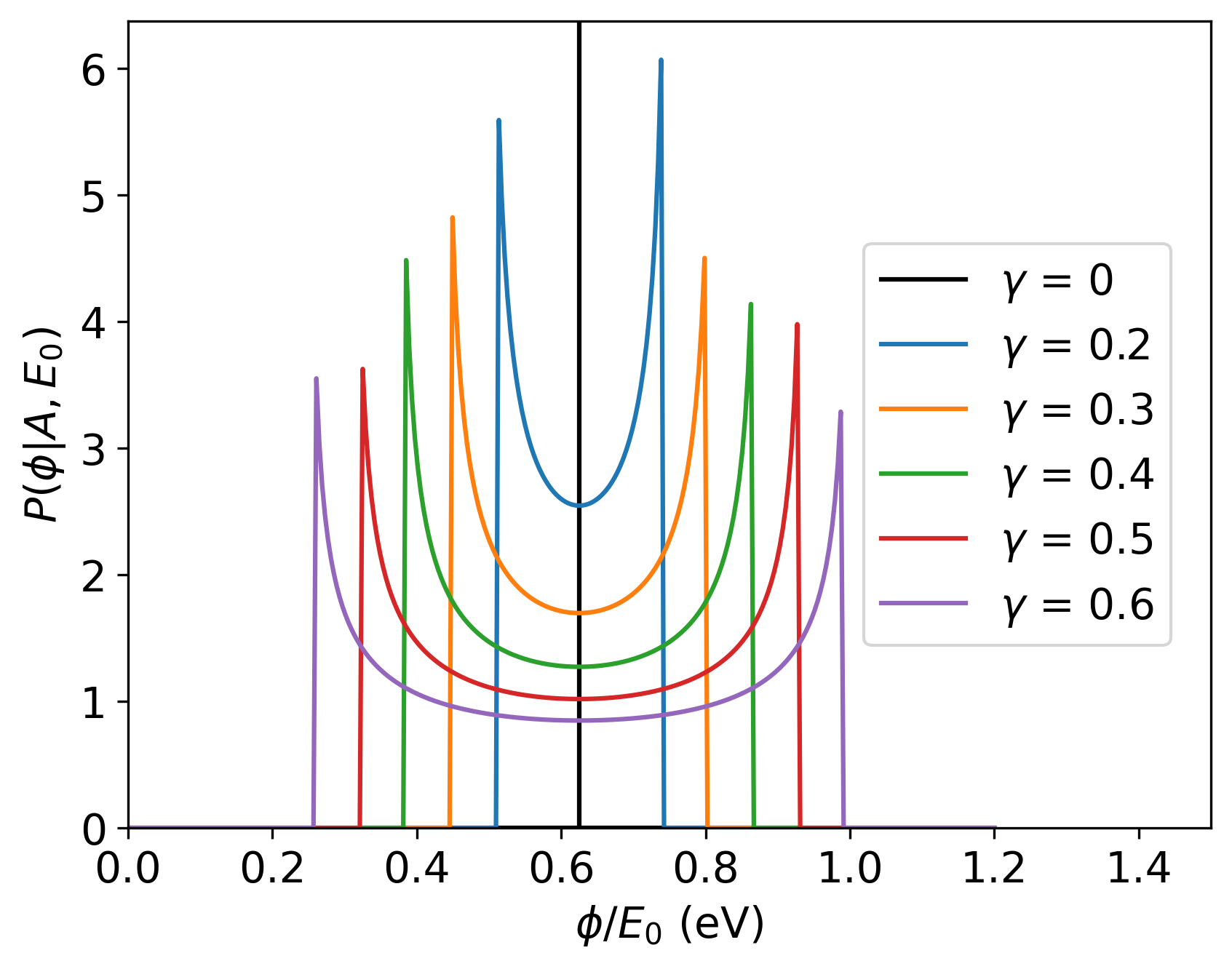}
\end{center}
\caption{Left, exact probability density $P(\phi|A, E_0)$ in \eqref{eq:probphi_exact} evaluated numerically for $N$ = 108. Right, asymptotic probability density $P(\phi|A, E_0)$ as given 
in \eqref{eq:probphi}.}
\label{fig:probphi_exact}
\end{figure}

The distribution in \eqref{eq:probphi_exact} does not seem to have a closed form but can be evaluated numerically, as shown in Fig.~\ref{fig:probphi_exact}. We can see curves for several values of 
$\gamma$ ranging from 0 to 0.6 on both the left and right panels. On the left panel, corresponding to the exact probability density in \eqref{eq:probphi_exact}, we can observe that the distribution has a 
Gaussian shape when no driving ($\gamma$ = 0) is considered. Two clear peaks and a valley develop when $\gamma$ is increased, and accordingly the variance of $\phi$ also increases. On the right side panel, 
showing the asymptotic probability density, two sharp peaks appear for non-zero values of $\gamma$, and again a deep valley forms between those peaks. Thus the behavior is an extreme form of the features 
observed for the exact density.

Despite the fact that \eqref{eq:probphi_exact} seems intractable, the moments of $P(\phi|A, E_0)$ can be computed in closed form using \eqref{eq:moments_E} together with the marginalization rule
\begin{equation}
\big<\phi^m\big>_{A, E_0} = \int_0^\infty dE P(E|A, E_0)\big<\phi^m\big>_E.
\end{equation}
 
\noindent
We obtain
\begin{equation}
\big<\phi^m\big>_{A, E_0} = \frac{\Gamma(\alpha+1)\Gamma(\alpha+m+1-\frac{3N}{2})}{\Gamma(\alpha+m+1)\Gamma(\alpha+1-\frac{3N}{2})}\,\big<E^m\big>_{A, E_0},
\end{equation}
which upon replacing \eqref{eq:moments} becomes
\begin{equation}
\big<\phi^m\big>_{A, E_0} = \phantom{.}_2F_1\Big(\frac{1-m}{2}, -\frac{m}{2}, 1; \gamma^2\Big)\cdot {E_0}^m \frac{\Gamma(\alpha+1)\Gamma(\alpha+m+1-\frac{3N}{2})}{\Gamma(\alpha+m+1)\Gamma(\alpha+1-\frac{3N}{2})},
\end{equation}
that is,
\begin{equation}
\label{eq:moments_osc}
\big<\phi^m\big>_{A, E_0} = \phantom{.}_2F_1\Big(\frac{1-m}{2}, -\frac{m}{2}, 1; \gamma^2\Big)\cdot \big<\phi^m\big>_{E_0}.
\end{equation}

Clearly we recover the microcanonical result for $\gamma$ = 0 because $\phantom{.}_2F_1(a, b, c; 0)$ = 1 for all $a, b, c$. Similarly, because $\phantom{.}_2F_1(0, b, c; z)$ = 1 for all 
$b, c, z$ it readily follows that
\begin{equation}
\label{eq:mean_osc}
\big<\phi\big>_{A, E_0} = \big<\phi\big>_{E_0} = E_0\left[1-\frac{3N}{2(\alpha+1)}\right].
\end{equation}

Here we can see that the mean potential energy is not affected by the amplitude of oscillation of the total energy. The relative variance of $\phi$ given $A$ and $E_0$ is
\begin{equation}
\label{eq:relvar_osc}
\frac{\big<(\delta \phi)^2\big>_{A, E_0}}{\big<\phi\big>_{A, E_0}^2}
= \frac{1}{2}\left(\frac{\alpha+1}{\alpha+2}\right)\left(\frac{\alpha + 2 - \frac{3N}{2}}{\alpha + 1 - \frac{3N}{2}}\right)\big(\gamma^2 + 2\big) - 1,
\end{equation}
and can see that \eqref{eq:relvar_osc} agrees with \eqref{eq:relvar_E} when $\gamma \rightarrow 0$. Moreover, it follows that the variance cannot decrease with respect to 
the one in \eqref{eq:relvar_E}, that is,
\begin{equation}
\big<(\delta \phi)^2\big>_{A, E_0} \geq \big<(\delta \phi)^2\big>_{E_0}.
\end{equation}

\noindent
The result in \eqref{eq:moments_osc}, as well as those in \eqref{eq:mean_osc} and \eqref{eq:relvar_osc} are exact for all sizes $N \geq 1$. The limit $N \rightarrow \infty$ of 
\eqref{eq:relvar_osc} yields
\begin{equation}
\lim_{N \rightarrow \infty} \frac{\big<(\delta \phi)^2\big>_{A, E_0}}{\big<\phi\big>_{A, E_0}^2} = \frac{\gamma^2}{2}.
\end{equation}

\noindent
Similarly, for $N \gg 1$, we can use Stirling's approximation to obtain
\begin{equation}
\label{eq:moments_approx}
\big<\phi^m\big>_{A, E_0} \approx (R E_0)^m \phantom{.}_2F_1\Big(\frac{1-m}{2}, -\frac{m}{2}, 1; \gamma^2\Big),
\end{equation}
with $R$ as defined in \eqref{eq:Rdef}, and this implies that the distribution of the dimensionless variable
\begin{equation}
z \defeq \frac{\phi}{RE_0}
\end{equation}
only depends on $\gamma$. Comparison of the approximate moments in \eqref{eq:moments_approx} with the exact moments \eqref{eq:moments} of the distribution in \eqref{eq:probE} tells 
us that $\phi/R$ follows the same distribution as $E$, therefore we can write
\begin{equation}
\label{eq:probphi}
P(\phi|A, E_0) \approx \frac{1}{\pi\,R E_0}\frac{\Theta\big(\gamma - |z-1|\big)}{\sqrt{z-(1-\gamma)}\sqrt{1 + \gamma -z}},
\end{equation}
and, in this approximation, $z$ is constrained to the interval
\begin{equation}
1-\gamma \leq z \leq 1+\gamma.
\end{equation}

\section{Molecular dynamics simulation}
\label{sec:md}

Computational simulations of a Lennard-Jones (LJ) crystal were performed to evaluate its potential energy distribution. Several different values of $A$ ranging from 0 to 2 were considered in order to 
establish a clear connection between the simulations and the distributions for inverse temperature and potential energy. The simulations were performed using a homemade software written in the Rust 
computational language~\cite{matsakis2014rust} for efficiency. For each value of $A$, several classical dynamics simulations were performed. The simulation consists of Ar atoms with a lattice constant 
$a=5.256$~\AA~in an FCC structure with 108 atoms. The interatomic potential used in the molecular dynamics simulations is the well-known Lennard-Jones pair potential, where
\begin{equation}
\Phi(\bm{r}_1, \ldots, \bm{r}_N) = \sum_{i=1}^N \sum_{j < i} V(|\bm{r}_i-\bm{r}_j|)
\end{equation}
with
\begin{equation}
V(r) = 4\epsilon \left[ \left( \frac{\sigma}{r} \right)^{12} - \left( \frac{\sigma}{r} \right)^{6} \right].
\end{equation}

Here $\epsilon$ is the depth of the potential well, $\sigma$ is the finite distance at which the inter-particle potential is zero, and $r$ is the distance between the particles. In our case, the parameters 
were used, namely $\epsilon/k_B = 120$ K, and $\sigma$ = 3.4~\AA~\cite{allen1989, Rapaport2004}. The potential energy of the ideal crystal, denoted by $\phi_0$, must be subtracted from the simulation data 
for consistency with the assumptions of our model, where zero temperature must correspond to zero total energy. The values of the properties defining the LJ system are summarized in Table~\ref{tbl:sysprop}. 

Table~\ref{tbl:run_params} shows the parameters used in some of the molecular dynamics runs and the values of the corresponding thermodynamical properties.

\begin{table}[h!]
\begin{center}
\begin{tabular}{|c|c|c|c|c|c|c|}
\hline
$N$ & $a$ (\AA) & $\sigma$ (\AA) & $\epsilon$ (K$\cdot k_B$) & $\phi_0$ (eV) & $\alpha/N$ & $\gamma_c$ \\
\hline
108 & 5.256 & 3.4 & 120 & -17.581 & 4.1652 & 0.0666413 \\
\hline
\end{tabular}
\end{center}
\caption{Properties of the LJ system used in molecular dynamics simulations.}
\label{tbl:sysprop}
\end{table}

\begin{table}[h!]
\begin{center}
\begin{tabular}{|c|c|c|c|c|c|c|}
\hline
Run & $A$ (eV) & $T_0$ (K) & $E_0$ (eV) & $\gamma$ & $T_S$ (K) & $T_S^{\dagger}$ (K) \\
\hline
509 & 0.5 & 50  & 2.1643 & 0.2303 & 54.33152 & 56.96488 \\
505 & 0.7 & 50  & 2.1411 & 0.3259 & 52.21707 & 54.79973 \\
501 & 1.1 & 50  & 2.1770 & 0.5037 & 48.51658 & 51.03723 \\
\hline
468 & 0.1 & 100 & 3.9725 & 0.0271 & 102.4402 & 103.5647 \\
447 & 0.3 & 100 & 3.7981 & 0.0792 & 97.67073 & 100.8323 \\
450 & 0.5 & 100 & 3.9976 & 0.1249 & 102.3180 & 105.4649 \\
454 & 0.7 & 100 & 3.9634 & 0.1762 & 100.6439 & 103.7880 \\
465 & 1.1 & 100 & 4.0459 & 0.2710 & 100.4638 & 103.6523 \\
514 & 1.7 & 100 & 4.4135 & 0.3840 & 105.1264 & 108.3611 \\
515 & 2.0 & 100 & 4.4395 & 0.4491 & 102.3233 & 105.5759 \\
\hline
511 & 0.5 & 150 & 6.6426 & 0.0755 & 170.8695 & 173.2787 \\
507 & 0.7 & 150 & 6.8160 & 0.1027 & 174.9016 & 177.2377 \\
503 & 1.1 & 150 & 5.8826 & 0.1865 & 149.0887 & 151.8897 \\
\hline
512 & 0.5 & 200 & 9.2725 & 0.0546 & 238.8443 & 240.1096 \\
508 & 0.7 & 200 & 7.8657 & 0.0891 & 202.1019 & 203.9980 \\
504 & 1.1 & 200 & 8.9802 & 0.1224 & 229.9184 & 231.3308 \\
\hline
\end{tabular}
\end{center}
\caption{Parameters used in several molecular dynamics simulation runs, together with measured properties. The value $T_S^{\dagger}$ is computed from the expression in 
\eqref{eq:beta_S_alt}.}
\label{tbl:run_params}
\end{table}

\begin{figure}[h!]
\begin{center}
\includegraphics[height=0.25\textheight]{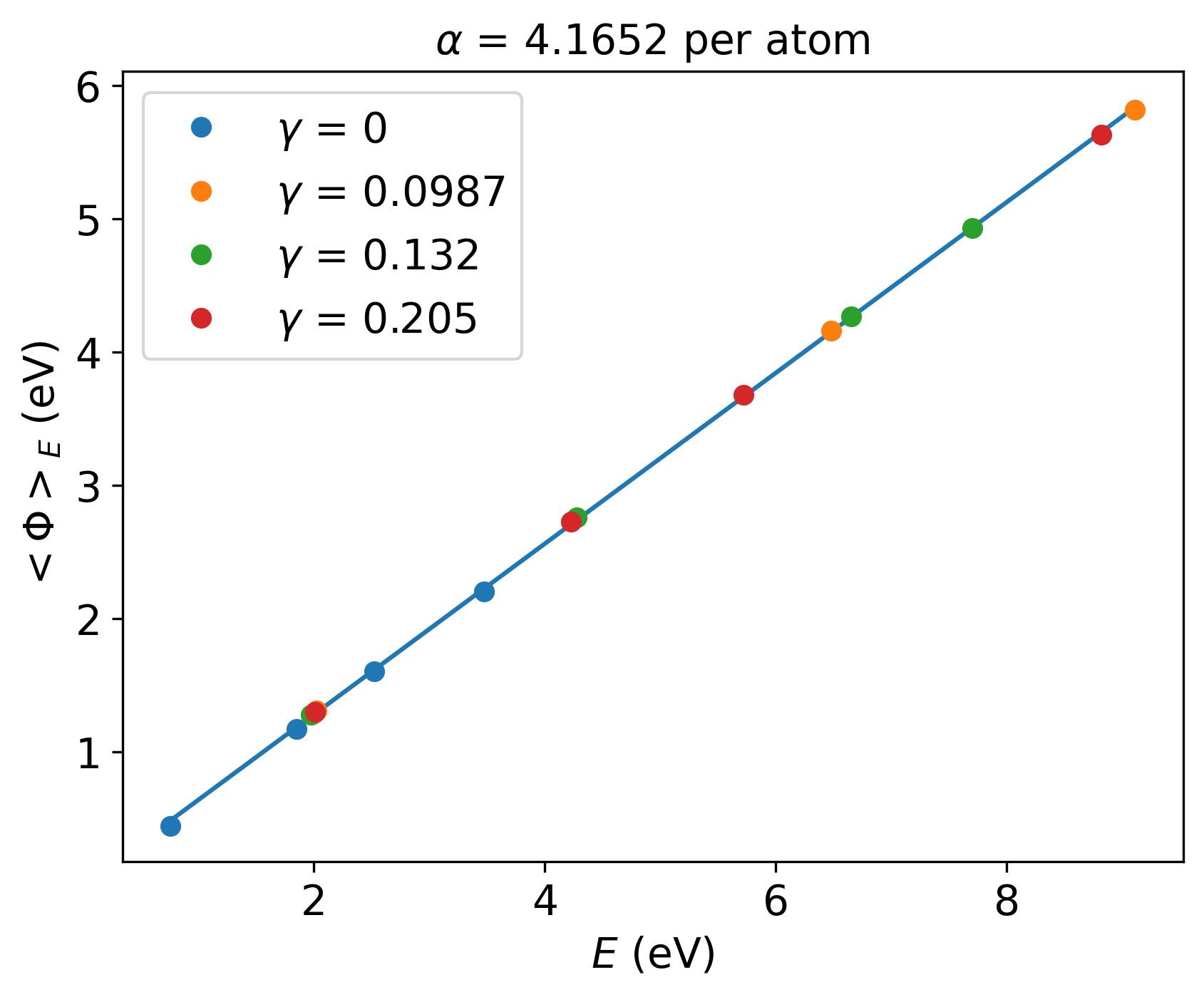}
\end{center}
\caption{Mean potential energy as a function of total energy for several values of $\gamma$, that is, corresponding to \eqref{eq:mean_osc}.}
\label{fig:caloric}
\end{figure}

Once the potential energy data is extracted from the simulations, we analyze its statistical distribution in order to compare them with the theoretical results of the superstatistical framework. 
Fig.~\ref{fig:caloric} shows the mean potential energy as a function of total energy for several values of $\gamma$. We confirm the linear relationship predicted by \eqref{eq:mean_phi_E} and the fact that 
the mean potential energy is independent of $\gamma$. From the slope of the line we have determined $\alpha$ = 4.1652 per atom.

\begin{figure}[b!]
\begin{center}
\includegraphics[height=0.25\textheight]{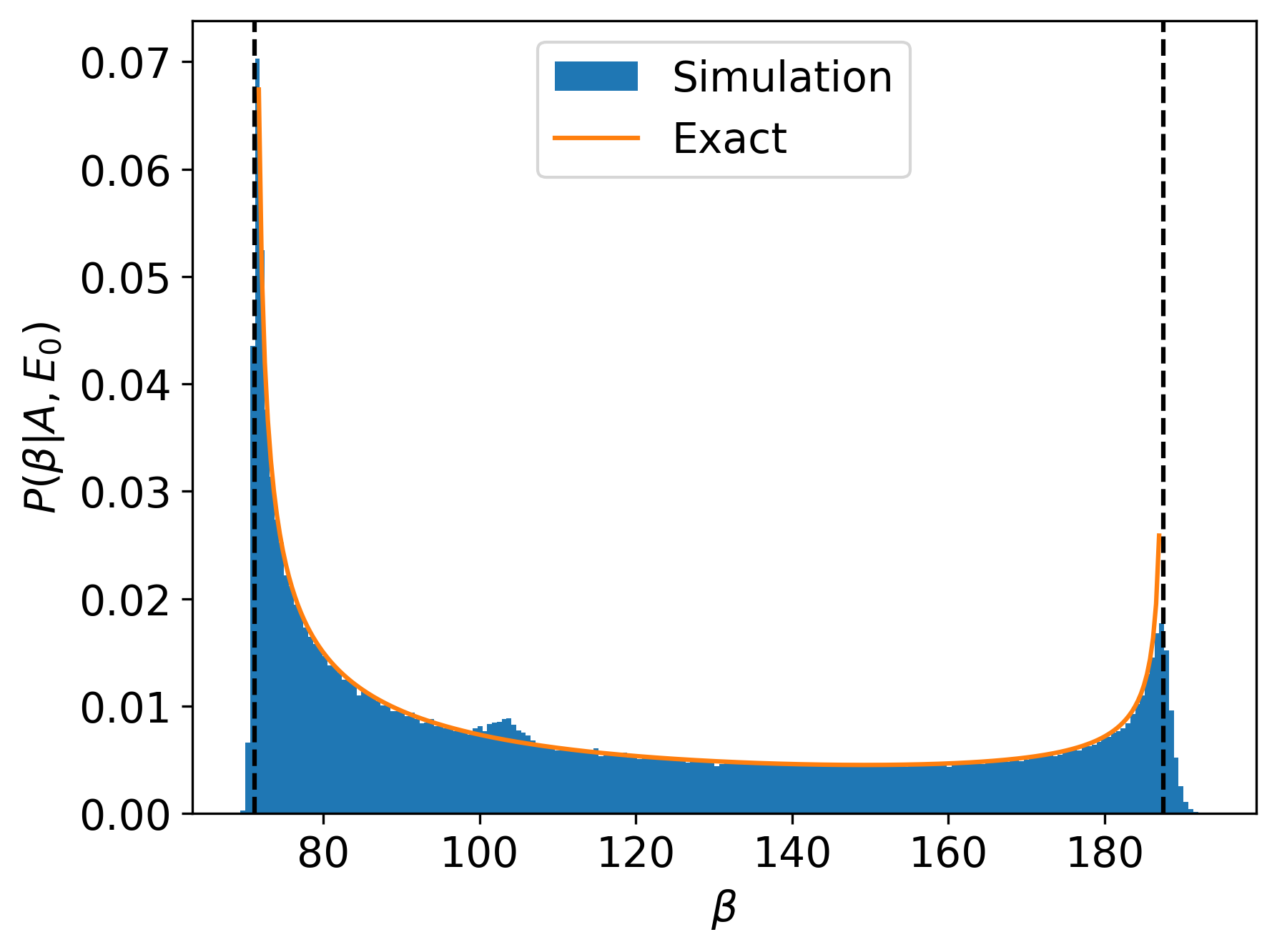}
\includegraphics[height=0.25\textheight]{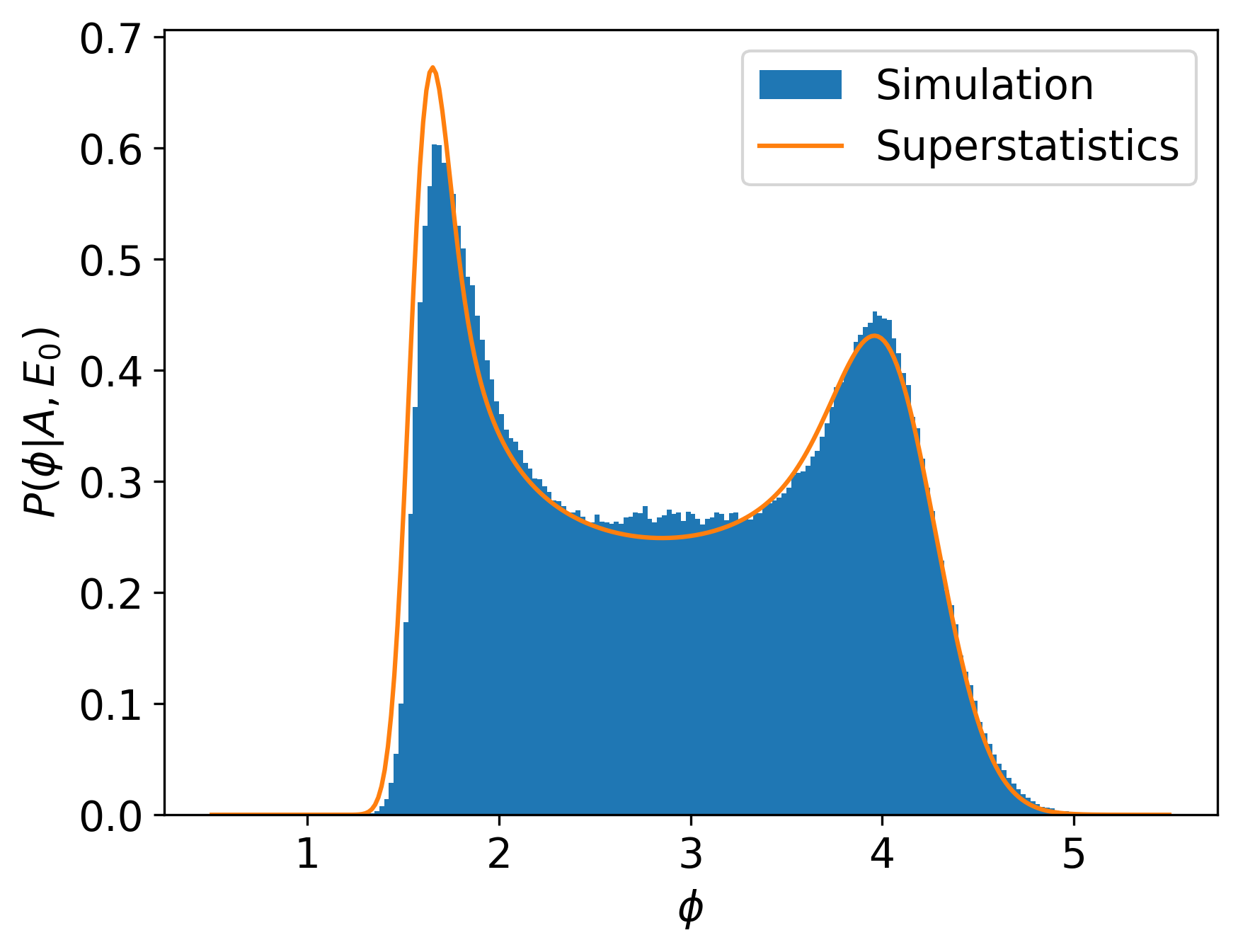}
\end{center}
\caption{Left, empirical distribution of inverse temperature $\beta_\Omega$ versus \eqref{eq:probbeta}. Right, superstatistical distribution of potential energies given by 
\eqref{eq:probphi_super} versus the empirical distribution from molecular dynamics simulation.}
\label{fig:superstat}
\end{figure}

\begin{figure}[h!]
\begin{center}
\includegraphics[height=0.25\textheight]{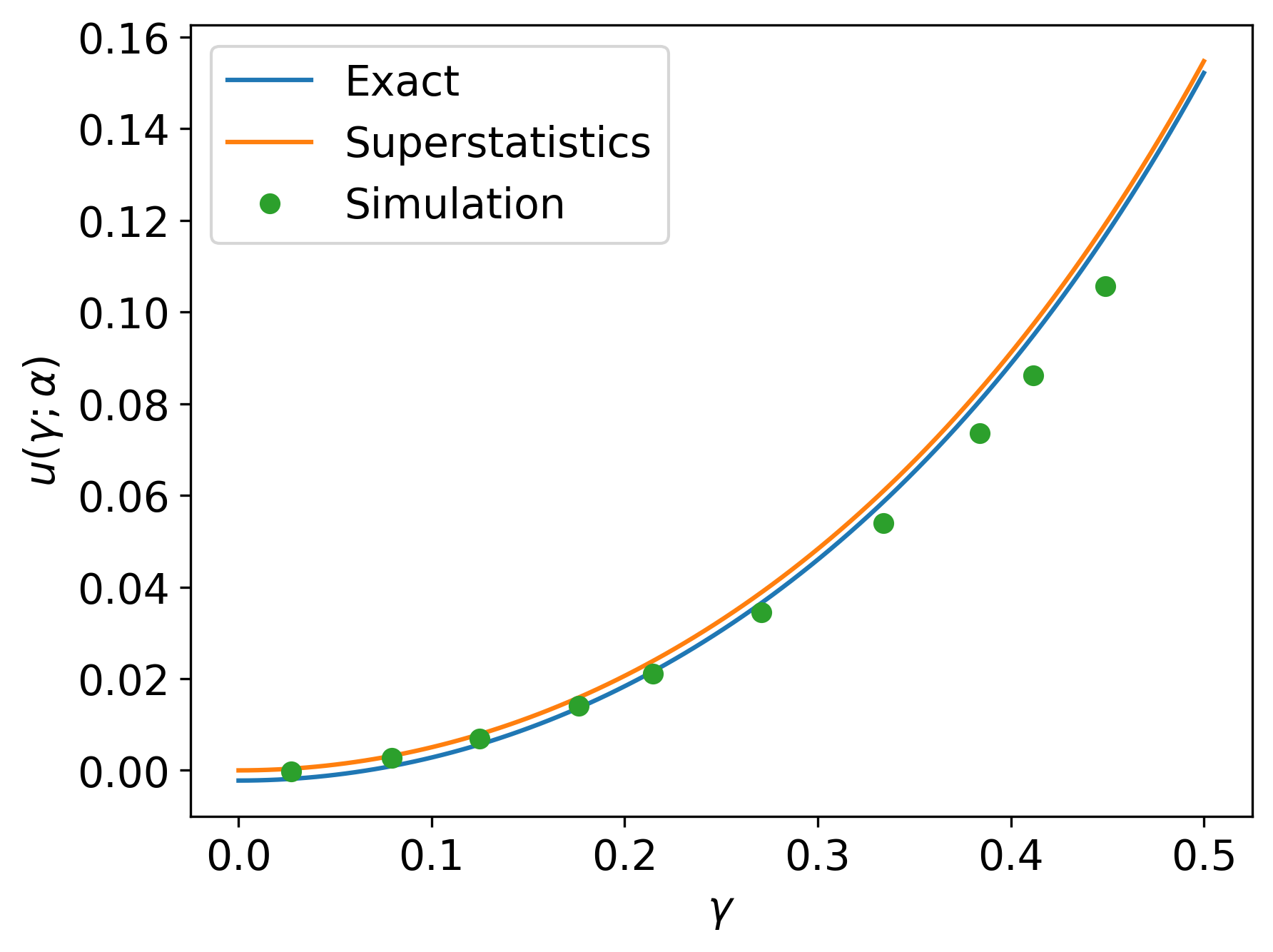}
\end{center}
\caption{Reduced inverse temperature correlation $u$ as a function of $\gamma$ for $N$ = 108 atoms and $\alpha/N$ = 4.1652. The blue curve is the exact prediction in \eqref{eq:upred}, 
while the orange curve is the asymptotic approximation in \eqref{eq:upred_asympt}. The green circles represent values computed from molecular dynamics simulations using \eqref{eq:umeas}.}
\label{fig:ugamma}
\end{figure}

Fig.~\ref{fig:superstat} compares the predictions of superstatistics with the molecular dynamics simulation. The left panel shows the agreement between the expected distribution of inverse temperatures 
and the empirical distribution from molecular dynamics. In the right panel, the superstatistical probability density for $\phi$ is compared with the empirical distribution from run 515 of 
Table~\ref{tbl:run_params}, and we see a precise agreement even for $N$ = 108 atoms. Similarly, Fig.~\ref{fig:ugamma} shows the relationship between $u$ and $\gamma$ as predicted in \eqref{eq:upred}, 
together with the superstatistical limit in \eqref{eq:super_relvar} and the values computed from molecular dynamics simulations.

\begin{figure}[h!]
\begin{center}
\includegraphics[height=0.25\textheight]{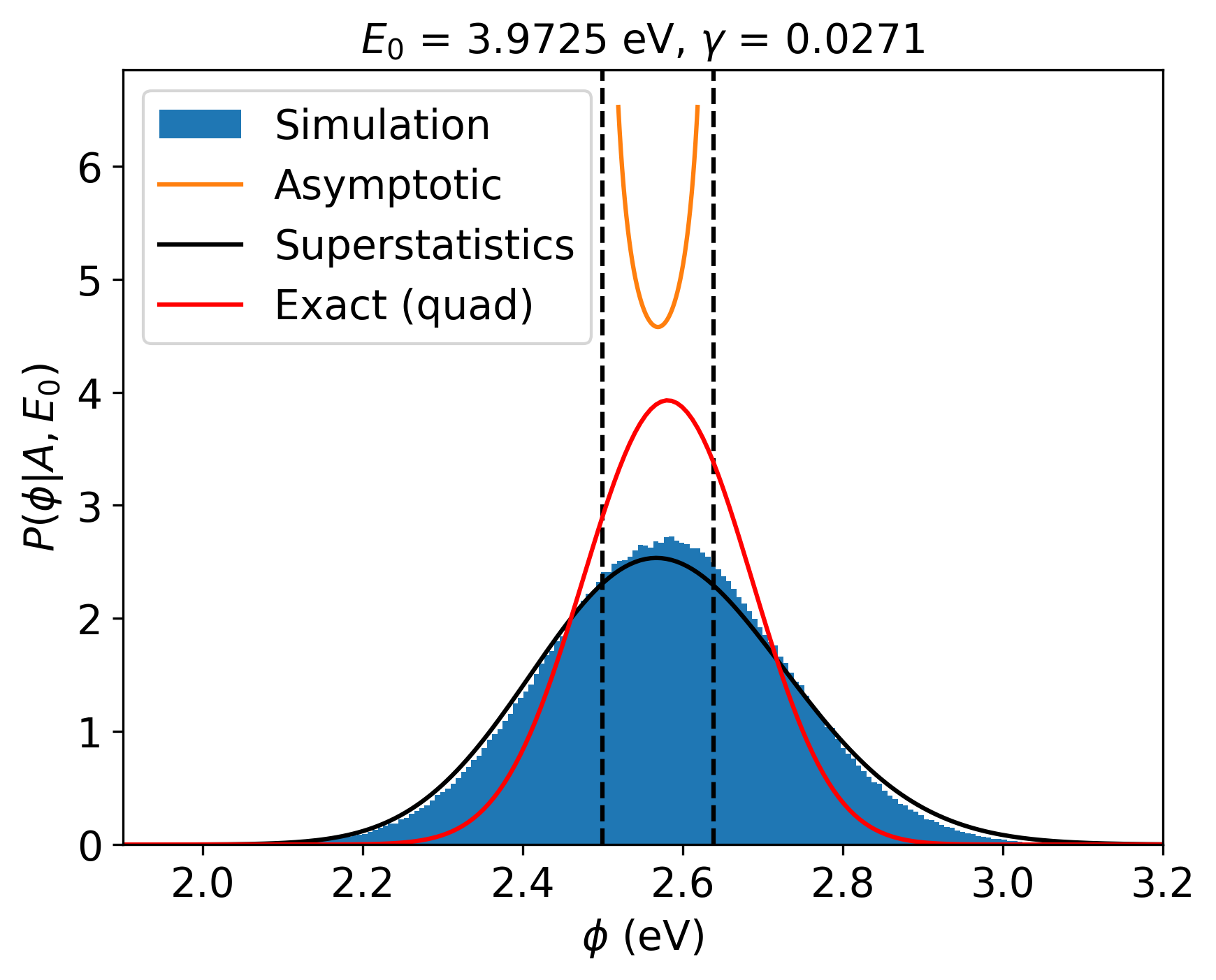}
\includegraphics[height=0.25\textheight]{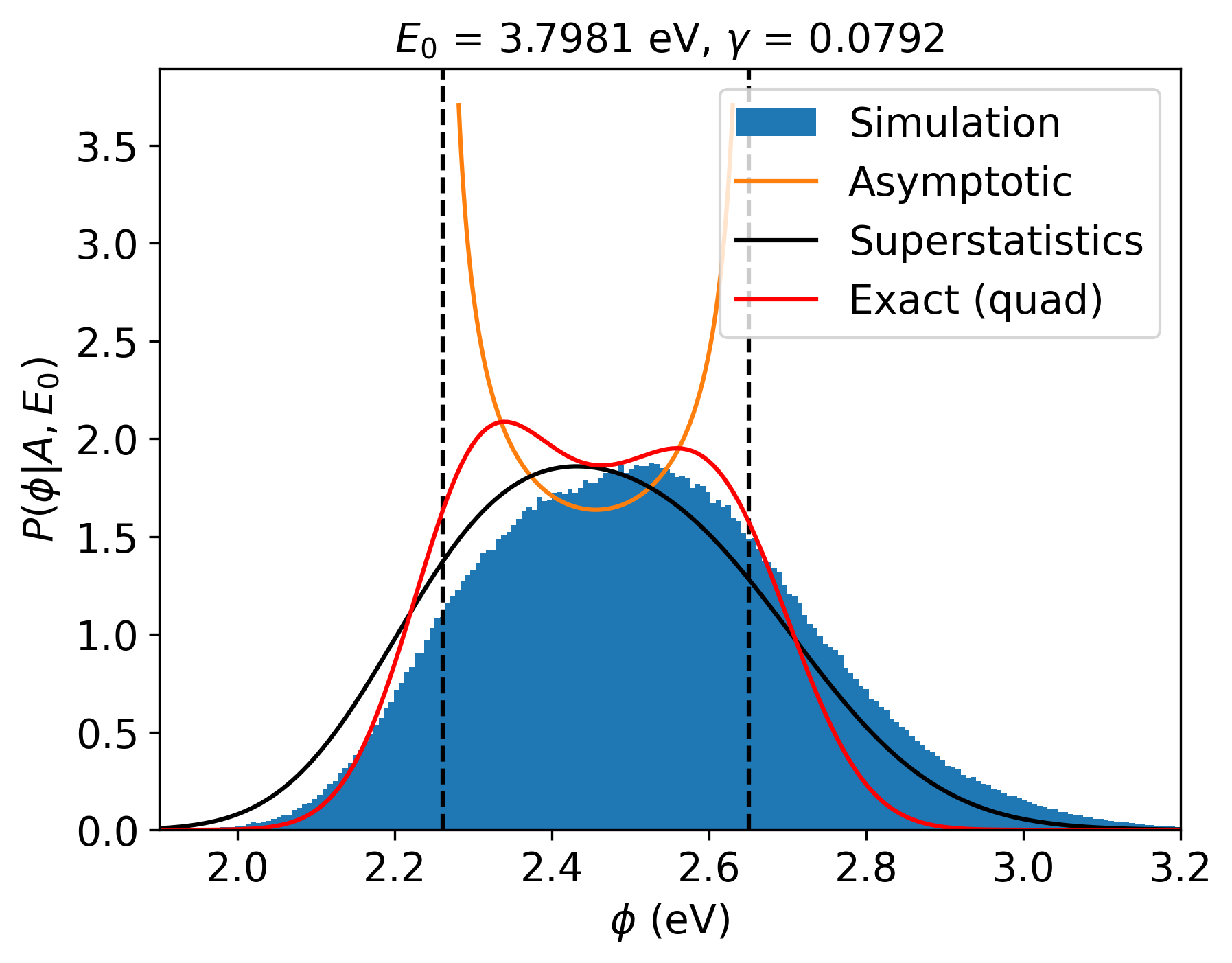}
\end{center}
\caption{Left, distribution of potential energies for $\gamma$ = 0.0271 and $E_0$ = 3.9725 eV, corresponding to run 468 in Table \ref{tbl:run_params}. Right, $\gamma$ = 0.0792 and $E_0$ = 3.7981 eV, 
corresponding to run 447. The orange curve is the asymptotic approximation in \eqref{eq:probphi}, while the red curve is the exact distribution in \eqref{eq:probphi_exact} evaluated numerically and the 
black curve is the superstatistical distribution in \eqref{eq:probphi_super}.}
\label{fig:pot1}
\end{figure}

\begin{figure}[t!]
\begin{center}
\includegraphics[height=0.25\textheight]{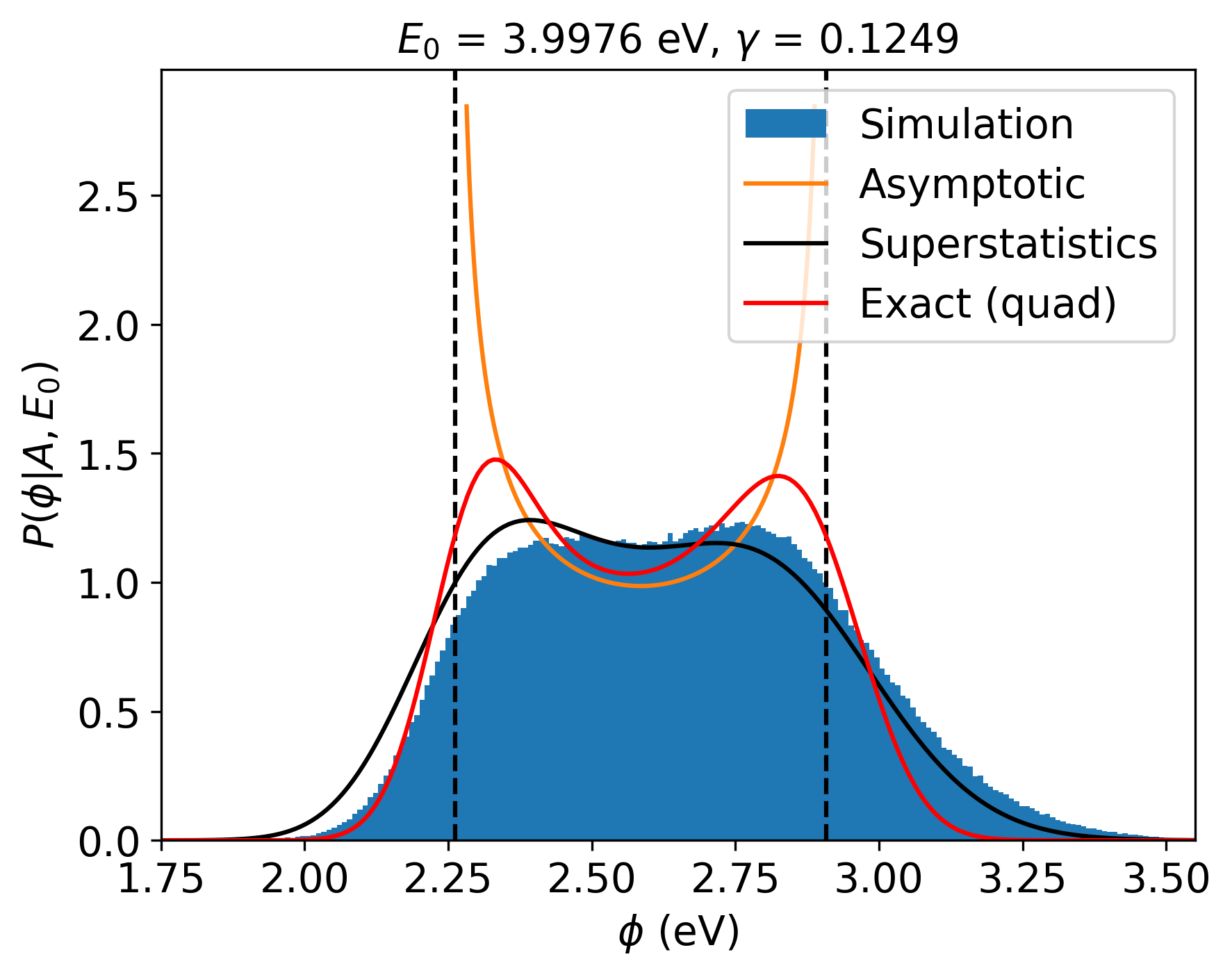}
\includegraphics[height=0.25\textheight]{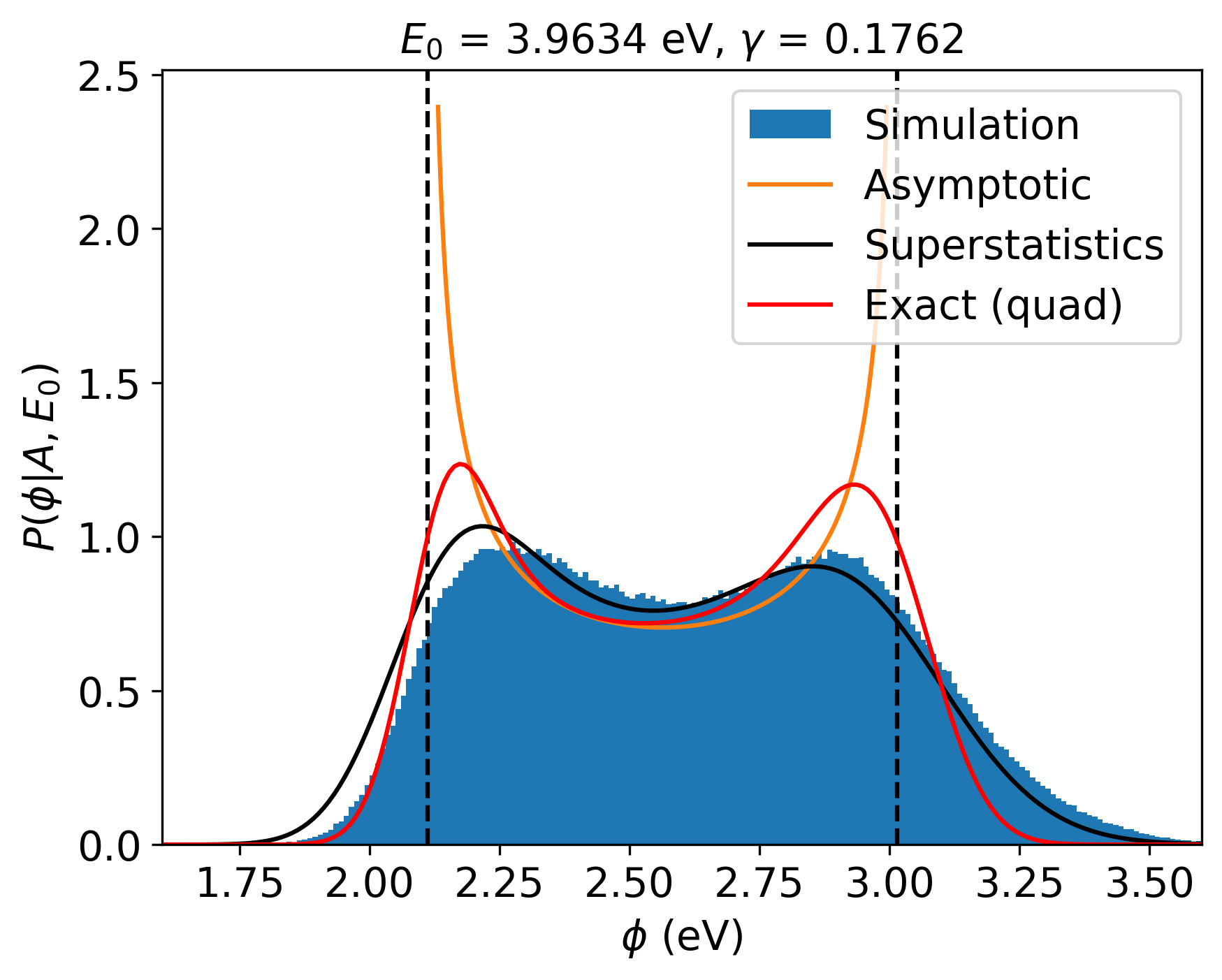}
\end{center}
\caption{Left, distribution of potential energies for $\gamma$ = 0.1249 and $E_0$ = 3.9976 eV, corresponding to run 450 in Table \ref{tbl:run_params}. Right, $\gamma$ = 0.1762 and $E_0$ = 3.9634 eV, 
corresponding to run 454. The orange curve is the asymptotic approximation in \eqref{eq:probphi}, while the red curve is the exact distribution in \eqref{eq:probphi_exact} evaluated numerically and the 
black curve is the superstatistical distribution in \eqref{eq:probphi_super}.}
\label{fig:pot2}
\end{figure}

\begin{figure}[t!]
\begin{center}
\includegraphics[height=0.25\textheight]{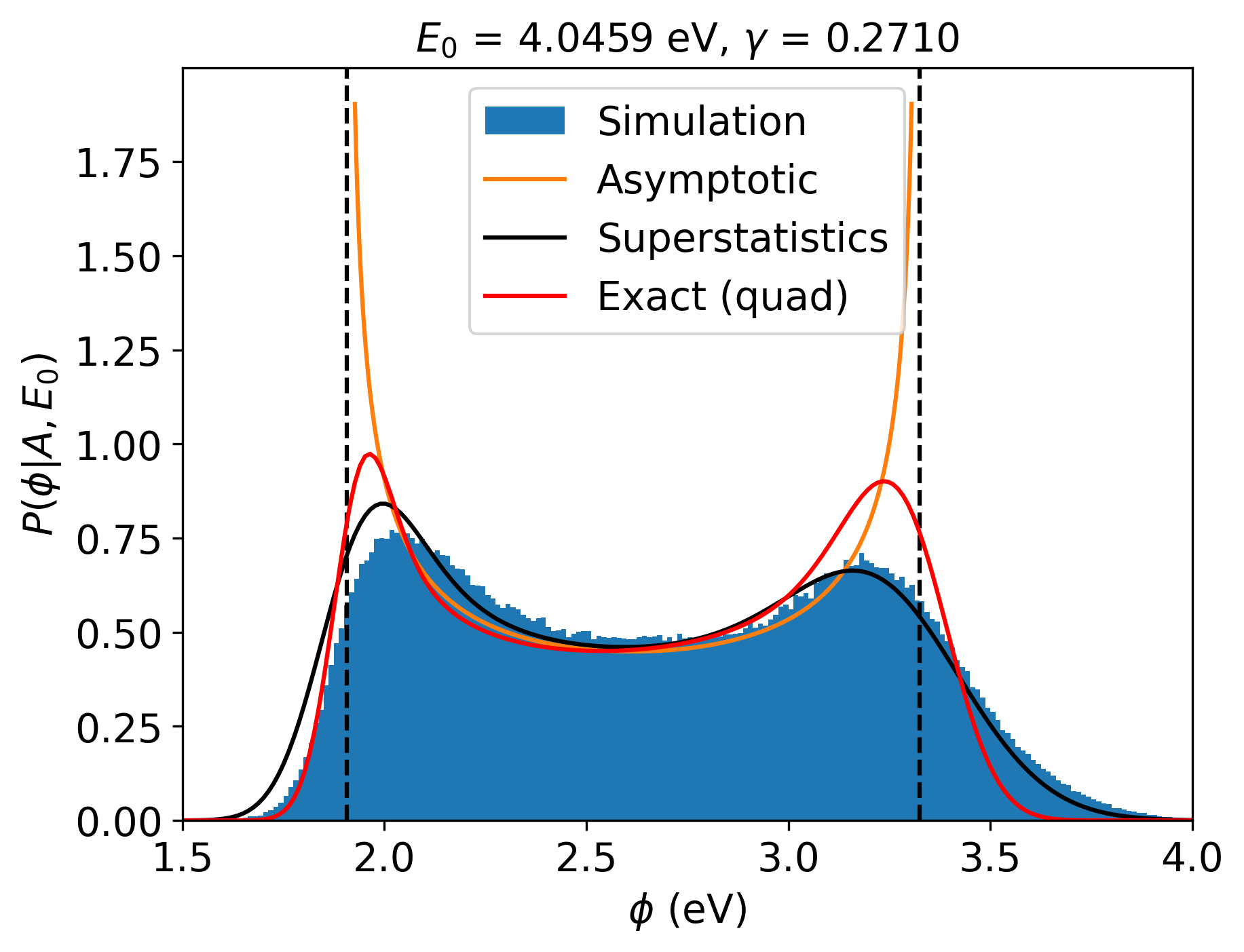}
\includegraphics[height=0.25\textheight]{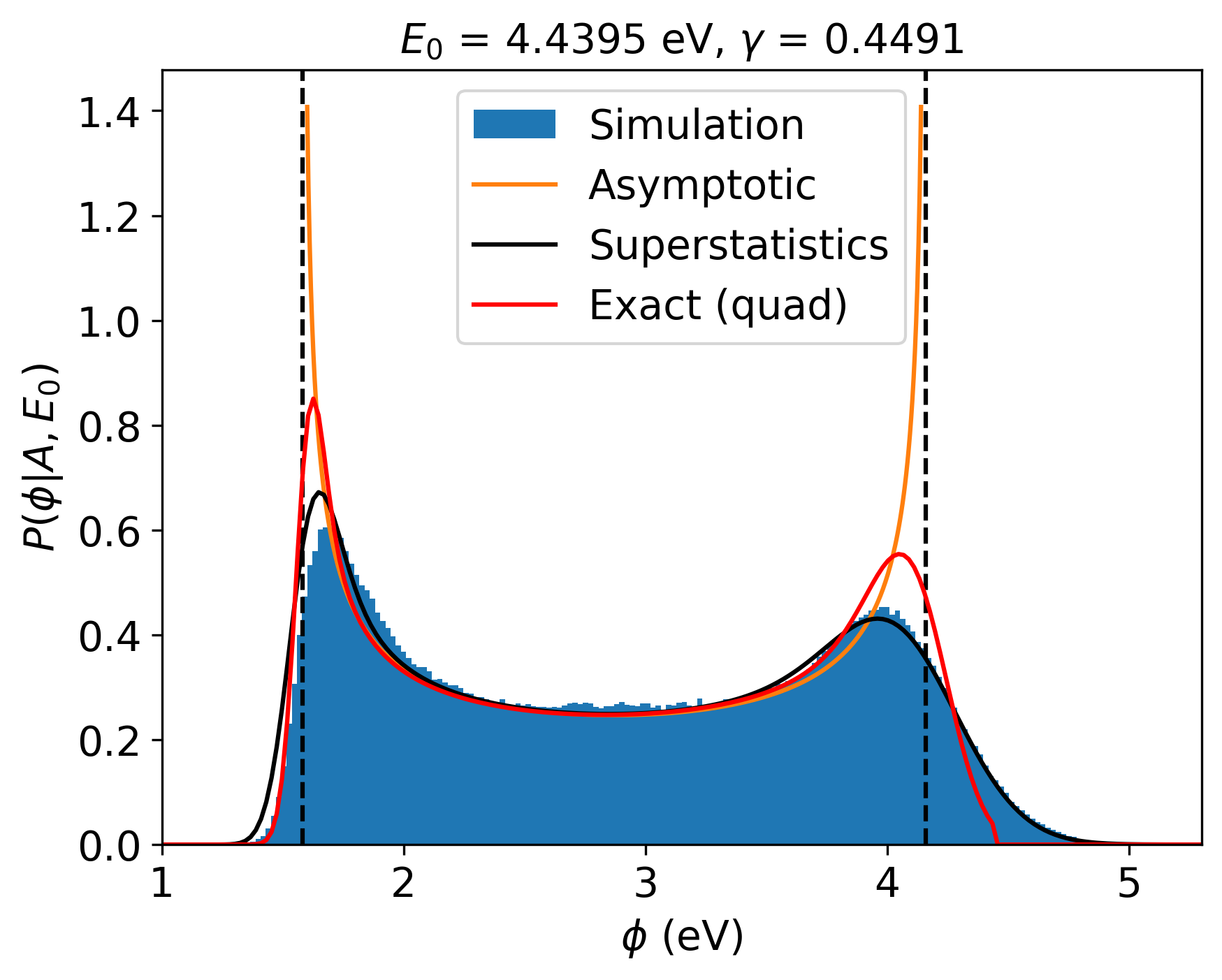}
\end{center}
\caption{Left, distribution of potential energies for $\gamma$ = 0.2710 and $E_0$ = 4.0459 eV, corresponding to run 465 in Table \ref{tbl:run_params}. Right, $\gamma$ = 0.4491 and $E_0$ = 4.4395 eV, 
corresponding to run 515. The orange curve is the asymptotic approximation in \eqref{eq:probphi}, while the red curve is the exact distribution in \eqref{eq:probphi_exact} evaluated numerically and the 
black curve is the superstatistical distribution in \eqref{eq:probphi_super}.}
\label{fig:pot3}
\end{figure}

Finally, Figs.~\ref{fig:pot1} to ~\ref{fig:pot3} depict the potential energy distributions observed in some of the molecular dynamics runs with parameters in Table~\ref{tbl:run_params}, together with the 
exact and superstatistical predictions. First, the left panel of Fig.~\ref{fig:pot1} shows the case of $\gamma$ = 0.0271 and $E_0$ = 3.9725 eV, corresponding to run 468. A Gaussian shape is observed here for 
both the simulation and the exact curve. On the right panel, we see the results for $\gamma$ = 0.0792 and $E_0$ = 3.7981 eV, corresponding to run 447. As $\gamma$ increases, the exact curve clearly defines 
two main peaks, which are moving to the boundaries of the asymptotic solution. For the case of simulation data, a more spread distribution is observed, and a clear peak can be easily defined, which suggests 
a transition of the distribution to a more complex one. Fig.~\ref{fig:pot2} presents two particular cases, in the left panel with $\gamma$ 0.1248 and $E_0$ = 3.9974 eV corresponding to run 450, while in the 
right panel we see the results from run 454, with $\gamma$ = 0.1762 and $E_0$ = 3.9632 eV. 

The left panel of Fig.~\ref{fig:pot3} shows the distribution of potential energies for $\gamma$ = 0.2710 and $E_0$ = 4.0459 eV, corresponding to run 465, while the right panel shows the results from 
run 515, having $\gamma$ = 0.4491 and $E_0$ = 4.4395 eV.


Here we can see that for high enough values of $\gamma$ the agreement between the simulation data and the predictions improves. Still we can notice significant deviations from the asymptotic 
approximation, particularly at the high--energy tails, where the exact and superstatistical distribution provide a more accurate description of the simulation data. This highlights the limitations of 
the asymptotic approximation in capturing the true behavior of potential energy distributions in finite-size systems. The exact numerical evaluation aligns closely with the simulation results, particularly 
at low energies, but not as expected at high energy, where an elongated tail is observed for the simulation data. Despite these, our results present a reasonable agreement, validating its effectiveness in 
modeling these systems.

\newpage
\section{Concluding remarks}
\label{sec:con}

In this study, we have successfully demonstrated the applicability of superstatistics to the thermodynamic behavior of finite, driven classical systems. By employing classical molecular dynamics simulations, 
we validated the theoretical framework of steady-state statistical mechanics and superstatistics for the description of the potential energy distributions observed under external energy fluctuations.

The agreement between the molecular dynamics simulations and the theoretical predictions clearly illustrates the robustness of the superstatistical approach. The analysis in terms of the inverse temperature 
covariance $\U$ and its dependence on the relative oscillation amplitude $\gamma$ provides information about the superstatistical regime. We confirm that the reduced inverse temperature covariance $u$ 
increases with $\gamma$ and for values exceeding a critical threshold $\gamma_c$, the system transitions to a regime where the superstatistical description becomes more pertinent. This finding is important 
in order to understand the conditions under which superstatistics is applicable even in the case of driven systems. In short, our results provide a fundamental connection between superstatistics and 
steady-state, finite-size thermodynamics, which opens the way for new applications of superstatistics in condensed matter physics.

\section*{Acknowledgments}

\noindent
Funding from ANID FONDECYT 1220651 is gratefully acknowledged. This computational work was supported by the NLHPC (ECM-02) and FENIX (UNAB) supercomputing infrastructures.

\appendix
\section{Configurational density of states}
\label{sec:appendix}

In the following, we will deduce the form in \eqref{eq:cdos} for the configurational density of states of a system with constant microcanonical heat capacity. First, 
consider the canonical partition function $Z(\beta)$ written in terms of the density of states $\Omega(E)$, that is,
\begin{equation}
\label{eq:zlaplace}
Z(\beta) = \int_0^\infty dE\Omega(E)\exp(-\beta E).
\end{equation}

As is well-known, for a Hamiltonian as in \eqref{eq:ham} the partition function is the product of a kinetic part, denoted by $Z_K(\beta)$ and a configurational part, 
$Z_\Phi(\beta)$, thus
\begin{equation}
Z(\beta) = Z_K(\beta)\,Z_\Phi(\beta)
\end{equation}
with 
\begin{equation}
\label{eq:zconf}
Z_\Phi(\beta) = \int_0^\infty d\phi\,\D(\phi)\exp(-\beta \phi)
\end{equation}
and where the kinetic part is given by
\begin{equation}
Z_K(\beta) = \int_0^\infty dK\,\Omega_K(K)\exp(-\beta K) = \int_0^\infty dK\,\big(W K^{\frac{3N}{2}-1}\big)\exp(-\beta K) = W\,\Gamma(3N/2)\beta^{-\frac{3N}{2}},
\end{equation}
after replacing \eqref{eq:kindos}. By replacing \eqref{eq:dos} into \eqref{eq:zlaplace} we readily have
\begin{equation}
Z(\beta) = \Omega_0\Gamma(\alpha+1)\beta^{-(\alpha+1)},
\end{equation}
and therefore the configurational part of the partition function is given by
\begin{equation}
\label{eq:zconf_val}
Z_\Phi(\beta) = \frac{Z(\beta)}{Z_K(\beta)} = \frac{\Omega_0\Gamma(\alpha+1)}{W\Gamma(3N/2)}\beta^{-\alpha-1+\frac{3N}{2}}.
\end{equation}

\noindent
Applying the inverse Laplace transform to \eqref{eq:zconf} and replacing \eqref{eq:zconf_val} we finally obtain
\begin{equation}
\D(\phi) = D_0\,\phi^{\alpha-\frac{3N}{2}}
\end{equation}
with
\begin{equation}
D_0 = \frac{\Omega_0\Gamma(\alpha+1)}{W\Gamma(3N/2)\Gamma(\alpha-3N/2+1)}.
\end{equation}

\bibliography{driven_lj}
\bibliographystyle{unsrt}

\end{document}